\documentclass{sig-alternate}
\usepackage[hyphens]{url}
\usepackage{eqnarray}
\usepackage{multirow}
\usepackage{graphicx}

\usepackage{amsmath}
\usepackage[export]{adjustbox}
\usepackage{listings}
\usepackage{amssymb}

\usepackage{subfigure}
\usepackage{balance}
\usepackage{xspace}
\usepackage{color,colortbl}

\usepackage{wasysym}

\lstdefinelanguage{json}{
    basicstyle=\normalfont\ttfamily,
    showstringspaces=false,
    breaklines=true,
}

\newcounter{note}

\newcommand{\hide}[1]{}
\newcommand{\ignore}[1]{}
\newcommand{\bitem}[1]{\item{\textbf{#1:}}}

\newcommand{\phish}[1][]{\ensuremath{\CIRCLE_{#1}}\xspace}
\newcommand{\legit}[1][]{\ensuremath{\Circle_{#1}}\xspace}
\newcommand{\prob}[1][]{\ensuremath{\LEFTcircle_{#1}}\xspace}

\newcommand{\phishclick}[1][]{\ensuremath{\overset{\curlywedge}{\CIRCLE_{#1}}}\xspace}
\newcommand{\legitclick}[1][]{\ensuremath{\overset{\curlywedge}{\Circle_{#1}}}\xspace}
\newcommand{\probclick}[1][]{\ensuremath{\overset{\curlywedge}{\LEFTcircle_{#1}}}\xspace}

\newcommand{\notphish}[1][]{\ensuremath{\overline{\phish[#1]}}\xspace}
\newcommand{\notlegit}[1][]{\ensuremath{\overline{\legit[#1]}}\xspace}
\newcommand{\notprob}[1][]{\ensuremath{\overline{\prob[#1]}}\xspace}

\newcommand{\eone}{\ensuremath{e_{1}}\xspace}
\newcommand{\etwo}{\ensuremath{e_{2}}\xspace}
\newcommand{\ethree}{\ensuremath{e_{3}}\xspace}
\newcommand{\trust}{\ensuremath{T}\xspace}

\newcommand{\ott}{OTT messaging applications\xspace}
\newcommand{\caller}{caller ID applications\xspace}
\newcommand{\Caller}{Caller ID applications\xspace}

\begin{document}
\title{Abusing Phone Numbers and Cross-Application Features for Crafting Targeted Attacks}

\numberofauthors{4}

\author{
\alignauthor
Srishti Gupta\\
       \affaddr{Indraprastha Institute of Information Technology}\\
       \affaddr{Delhi}\\
       \email{srishtig@iiitd.ac.in}
\alignauthor
Payas Gupta\\
       \affaddr{New York University}\\
        \affaddr{Abu Dhabi}\\
       \email{payasgupta@nyu.edu}
\alignauthor Mustaque Ahamad\
       \affaddr{Georgia Institute of Technology}\\
       \affaddr{New York University Abu Dhabi}\\
       \email{mustaq@cc.gatech.edu}
\and \alignauthor Ponnurangam Kumaraguru\\
        \affaddr{Indraprastha Institute of Information Technology}\\
       \affaddr{Delhi}\\
       \email{pk@iiitd.ac.in}
}

\maketitle

\begin{abstract}
With the convergence of Internet and telephony, new applications (e.g., WhatsApp) have emerged as
an important means of communication for billions of users.
These applications are becoming an attractive medium for attackers to deliver 
spam and carry out more targeted attacks. Since such applications rely on phone numbers,
we explore the \textit{feasibility}, \textit{automation}, and 
\textit{scalability} of phishing attacks that can be carried out by abusing a phone number. 
We demonstrate a novel system that takes a potential victim's phone number as an input, 
leverages information from applications like Truecaller and Facebook about 
the victim and his / her social network, 
checks the presence of phone number's owner (victim) 
on the attack channels (over-the-top or \ott, voice, e-mail, or SMS),
and finally targets the victim on the chosen channel. 
As a proof of concept, we enumerate through a random pool of 1.16 million 
phone numbers. By using information provided by popular applications,
we show that social and spear phishing attacks can be launched against 51,409 and 180,000 users respectively. Furthermore, voice phishing or vishing attacks can be launched against 722,696 users. We also found 91,487 highly attractive targets who can be attacked by crafting whaling attacks.
We show the effectiveness of one of these attacks, phishing, by conducting an online roleplay user study. 
We found that social (69.2\%) and spear (54.3\%) 
phishing attacks are 
more successful than non-targeted phishing attacks (35.5\%) on \ott.
Although similar results were found for other mediums like e-mail, 
we demonstrate that due to the significantly increased user engagement 
via new communication applications and the ease with which phone numbers allow 
collection of information necessary for these attacks, there is a clear need for better protection 
of \ott. We propose some recommendations in this direction.

\end{abstract}
\keywords{Phone number, Over-The-Top, Phishing, Roleplay, Facebook, Truecaller, WhatsApp, Vishing.

\section{Introduction} \label{sec:intro}
%
%
We are being constantly targeted by cyber criminals who rely on a variety of online attacks to victimize
users and enterprises. Phishing, which is a form of social engineering attacks, is often used by such criminals to fraudulently gain
access to sensitive information or systems by impersonating a trusted party~\cite{jagatic2007social}. In the past, phishing attacks have
used the e-mail and web channels to reach their victims. However, recently,
there has been a tremendous growth of similar phishing
attempts over the telephony channel. New forms of phishing
attacks have emerged 
exploiting traditional text messaging services, i.e., SMS (smishing~\cite{smishing}) and voice phishing (vishing~\cite{vishing}).

Several factors make the telephony channel attractive for cyber criminals.
The convergence of telephony with the Internet has resulted in an 
unprecedented growth of new forms of online communication, 
especially mobile communication due to the advent of Over-The-Top (OTT) messaging 
applications (like WhatsApp, Viber, and WeChat~\cite{social2015}).
Because of the growing popularity of \ott, 
particularly WhatsApp, malicious actors are now abusing it for illicit activities
like delivering spam and phishing messages. 
Unsolicited messages like investment advertisements, adult conversation ads (random contacts requests) were seen to propagate on the channel in early 2015~\cite{adaptive-whatsapp}.

Vishing attacks are also increasing due to 1) low mobile-mobile calling plans with the 
advent of services like Skype and Google Voice. This allows spammers to pump 
out huge volumes of voice calls at a marginal cost, and 
2) easy caller ID spoofing due to Voice over IP phone technology that 
allows spammers to pick an area code and even the prefix number they 
want when they set up a new phone number. These numbers can be used to 
disguise where calls originate. 
To avoid falling victim to such vishing attacks
and know more about the incoming phone number, cloud-based
caller identification services are emerging to help in
getting additional information about the caller. 
Millions of people are using such applications, 
namely Truecaller~\cite{truecaller}, Facebook's Hello~\cite{hello}, and Whitepages \Caller and Block~\cite{current_caller_id}.

\ott and \caller use a phone (mobile) number, a personally identifiable piece of information 
with which an individual can be associated  uniquely, 
in most cases~\cite{zheleva2011privacy}. \ott use it to uniquely identify 
users and allow them to find their friends who also use the same application
and \caller
provide additional information about the calling phone number.
Prior research has shown that other Internet resources like
e-mail addresses can be exploited as an identifier
to launch targeted phishing attacks~\cite{jagatic2007social, sheng:who-falls-for-phish:-a-de:2010:lrfkq}, 
distributed phishing attacks~\cite{jakobsson2005distributed}, and to correlate user identities across social networking platforms~\cite{balduzzi2010abusing}. 
In this paper we demonstrate how attackers 
can exploit phone number as a
unique identifier and use cross-application features for
launching spear~\cite{sheng:who-falls-for-phish:-a-de:2010:lrfkq} and social 
phishing attacks~\cite{jagatic2007social}. 

A challenge spammers face when using an e-mail address is that this medium is heavily defended
and spammers often cannot ensure that a spam message has been delivered and seen by the target user.
Furthermore, unlike e-mail addresses that come from an unlimited pool and can be freely created,
phone numbers are a limited and controlled resource. People \emph{generally}
retain the same number for a long period due to the cost associated with it~\cite{sim-cost}. 
Also, phone numbers are a finite pool with a defined numbering plan. 
They can be easily enumerated by looping through the entire pool of number space. 
However, it is possible that not all phone numbers are currently 
allocated to users and some of 
them may be unassigned. Thus, determining if a phone number is 
currently assigned and its owner can be 
reached is a challenge that needs to be overcome 
before such applications can be targeted.



In this paper, we demonstrate
how a phone number can be used across multiple applications
to aggregate private and personal information about the owner of the phone number.
Such information can then be used for targeted attacks. 
For example, reverse-lookup 
contact feature used by \caller like Truecaller~\cite{truecaller} can be exploited
to find more details (name) about the phone number's owner.
Moreover, by correlating this with the public information present on online social networking platforms (e.g., Facebook), 
attackers can  determine the social circle (friends) of the victim. Such information can then be used to launch a variety of phishing attacks.


We focus on exploring how cross-application features can be exploited to harvest information that can 
facilitate targeted and non-targeted attacks on different channels viz., \ott, voice, e-mail, or SMS.
First, we demonstrate how to craft targeted spear and social
phishing attacks against a random pool of phone numbers on \ott. 
\ott allow attackers to find relevant 
information about targets by exploiting 
address book syncing feature which helps to discover friends on a 
given OTT messaging application (e.g., WhatsApp). 
Second, we demonstrate a novel targeted vishing attack that can be carried out by 
compromising the integrity of \caller; an attacker can create a convincing profile to gain victim's trust. 
Also, because the information (e.g. name) provided to these applications during registration is not verified fully, it is 
relatively easy to add accounts with false information and impersonate trusted entities such as banks.
By making a call appear to come from such entities, it is easy to
deceive people into giving out their personal information 
like bank account number, credit card number etc. 
The success rate of vishing attacks can be increased by making them more personalized
and targeted by collecting information about the victim from Truecaller and Facebook.
Third, we provide early evidence of the feasibility of crafting whaling attacks~\cite{whaling}, attacks that are targeted against the owners of 
vanity numbers~\cite{vanity}, phone numbers generally owned by people with high influence or high-net-worth individuals, who are very
attractive targets for criminals.

By developing 
an automated and scalable system that allows such attacks to be crafted at scale 
and evaluating the effectiveness of phishing attacks on \ott (as a proof of concept) with a roleplay user study, we make the following contributions on three different fronts:

\begin{itemize}
\bitem{Feasibility} 
This is the \emph{first} attempt to systematically understand the threat posed by the ease of correlating user information
across caller ID lookup application (Truecaller), and 
social networking application (Facebook). This was executed using
phone numbers as unique identifiers.
We show the attack is feasible with 
easily available computational resources, and poses 
a significant security and privacy threat.
An attacker can use these cross-application features to launch 
highly targeted attacks on multiple channels like \ott, voice, e-mail, or SMS.

\bitem{Automation} 
We design and implement an automated system that takes a phone number as an input, 
collects necessary information about the victim (owner of the phone number). It can 
automatically determine the target attack channel (\ott, voice, e-mail, or SMS), and 
finally crafts an attack vector 
(both targeted and non-targeted) to launch an attack against the victim on the chosen channel.

\bitem{Scalability} 
For 1,162,696 random pool of Indian phone numbers that we enumerated, it is possible to 
launch social and spear phishing attacks against 51,409 and 180,000 users respectively.
Vishing attacks can be launched against 722,696 users. We also found 91,487 highly influential victims who can be attacked by crafting whaling attacks against them.
\end{itemize}

To demonstrate the \emph{effectiveness} of phishing attacks on \ott, we present results by conducting a roleplay user study 
with 314 participants recruited from 
Amazon Mechanical Turk (MTurk). 
Our results are consistent with prior research on e-mail and online social networks~\cite{balduzzi2010abusing, jagatic2007social} 
and confirm empirically 
that social phishing (69.2\%) is the most successful attack on \ott, as compared to 
spear (54.3\%) and non-targeted phishing (35.5\%) attacks. 

%

To the best of our knowledge, this is the first exploration of large-scale targeted 
attacks abusing phone numbers along with an evaluation of the effectiveness of the 
attacks. 
Given that telephony medium is not as well defended as e-mail, we believe that these 
contributions offer a promising new direction and demonstrate the urgent need for 
better security for such applications.


%

\section{Related Work} \label{sec:related}
In this section, we briefly outline some of the prior research related to abusing phone numbers; vishing attacks and Spam over Internet Telephony (see Section~\ref{sec:related:vishing}) and abusing address book syncing feature of smartphones 
for user profiling (see Section~\ref{sec:related:abusebook}),
We also discuss some of the related work on  launching 
targeted attacks on 
online social media (see Section~\ref{sec:related:targetattacks}). 

\subsection{Vishing Attacks and SPIT} \label{sec:related:vishing}
Due to low cost and scalability of VoIP based calling systems, 
scammers are using the telephony channel to make millions of call and expand the vishing ecosystem. 
Prior work has explored the detection and ways to combat scams on VoIP. 
Griffin et al. demonstrated that vishing attacks can be carried out using VoIP~\cite{griffin2008vishing}. 
They illustrated how several vishing attacks can be crafted in order to increase information security awareness.
Chiappetta et al. analyzed VoIP CDRs (Call Detail Records) to build 
features that can classify normal or malicious users during voice communication~\cite{chiappetta2013anomaly}. 
The features were built using mutual interactions and communication patterns between the users.
Past literature demonstrates detection of spam over VoIP through semi-supervised clustering~\cite{wu2009spam}, 
constructing multi-stage spam filter based on trust and reputation of callers~\cite{dantu2005detecting}, 
comparing human communication patterns with hidden Turing tests to detect botnets~\cite{quittek2007detecting}, 
building a system using features like call duration, social networks, and global reputation~\cite{balasubramaniyan2007callrank}, 
proposing protection model based on user-profile framework such as users' habits~\cite{scat2011user}, 
placing telephone honeypots to collect intelligence about telephony attacks~\cite{payas_m3aawg, gupta2015phoneypot}, 
and using call duration and traffic rate~\cite{kim2009devs}.
Caller ID spoofing is being used by scammers to hide their real identity and make fraudulent calls. 
Researchers have implemented various solutions to detect caller ID spoofing, 
using covert channels built on timing estimation and call status for verification~\cite{mustafa2014you}, 
identifying the caller by tracing the calls to the corresponding SIP-ISUP interworking gateway~\cite{song2014ivisher},
using customer's phonebook feature for storing white and black lists for filtering unwanted voice calls~\cite{cai2005phonebook}, 
and detecting audio codecs in call path, calculating packet loss and noise profiles to determine source and path of the call~\cite{balasubramaniyan2010pindr0p}.

In this work, we present first evidence of feasibility of targeted vishing attacks by exploiting the integrity of the information provided by \caller e.g., Truecaller.

\subsection{Abusing Address Book Syncing in OTT Messaging Applications} \label{sec:related:abusebook}
Recent work shows that collection of user profiles can be automated and 
yields a lot of personal information like phone numbers, display names, and 
profile pictures~\cite{cheng2013bind,kim2015ve}.
Schrittwieser et al. analyzed popular \ott like 
WhatsApp, Viber, Tango etc. and evaluated their security models with a focus on 
authentication mechanisms~\cite{schrittwieser2012guess}. 
They also highlighted the enumeration and privacy-related attacks that are possible 
due to address book syncing feature of these applications. Antonatos et al. 
proposed HoneyBuddy, an active honeypot infrastructure designed to detect 
malicious activities in Instant Messaging applications like MSN~\cite{antonatos2010systematic}. 
It automatically 
finds people using a particular messaging service and adds them to its contacts. These
findings confirmed the ineffectiveness of existing security measures in Instant 
Messaging services. 

In our research, we further demonstrate how cross-application features can be abused to launch targeted attacks; address book sync feature in \ott and exploiting integrity of information in \caller.

\subsection{Targeted Attacks and User Profiling} \label{sec:related:targetattacks}
Bilge et al. launched automated identity theft attacks via profiling users on 
SNS (Social Networking Services) by employing friend relationship with the victims~\cite{bilge2009all}. 
The authors showed that people tend to accept friend requests from 
strangers on social networks. Balduzzi et al. presented experiments conducted on ``social phishing''~\cite{balduzzi2010abusing}. They crawled 
social networking sites to obtain publicly available information about users 
and manually crafted phishing e-mails containing certain information about them. 
This study showed that victims are more likely to fall for phishing 
attempts if some information about their friends or about themselves 
is included in the phishing e-mail. Jagatic et al. showed that Internet 
users might be over \textit{four times} more likely to become victims if 
the sender is an acquaintance~\cite{jagatic2007social}. 
Gupta et al. showed that 
inference attacks can be employed to harvest real interests of people 
and subsequently break mechanisms that use 
such personal information for user authentication~\cite{Gupta:2013:YLP:2484313.2484319}.
Huber et al. presented 
friend-in-the-middle-attack on Facebook which could leverage social information 
about users in an automated fashion~\cite{huber2011friend}. They further pointed out the possibility 
of context-aware spam and social phishing attacks, where attacks were found 
to be cheap in terms of cost and hardware. 
Boshmaf et al. highlighted vulnerabilities that can be exploited by social bots to infiltrate OSNs~\cite{boshmaf2013design}. They showed that social bots can mimic real users and exploit 
friendship network leading to strong privacy implications.
Kurowski showed a manual attack on WhatsApp to retrieve personal information 
about victims and proposed the feasibility of social phishing attacks 
against victims~\cite{kurowskiusing}. 

In this paper, 
we demonstrate how phone numbers can be used for \emph{automated} 
cross-application targeted attacks 
where people are attacked on one application by leveraging information 
from multiple other applications i.e. using Truecaller and Facebook to launch targeted phishing, vishing, and whaling attacks on \ott, voice, e-mail, or SMS.

\section{System Overview: Feasibility and Automation} \label{sec:approach}
In this section, we demonstrate the \emph{feasibility} and the \emph{ease} with which 
different targeted attacks can be crafted by abusing phone numbers. To \emph{automate} the whole process,
we build a system that exploits cross-application features to collect information about a phone's user 
and
determines the attack channel (\ott, voice, e-mail, or SMS) and  targeted or non-targeted attack vectors (see Figure~\ref{fig:system}). Specifically, the system 
has four main steps. a) Based on a numbering plan, phone numbers are randomly generated and inserted 
into an address book of a smartphone. This address book is on a device that is under the control of the attacker.
b) The system
fetches data from Truecaller and Facebook applications to determine any 
additional information about the owners of those phone numbers. 
c) After the information is aggregated, 
the system determines the attack channel viz. \ott, voice, e-mail, or SMS, and d) finally crafts an attack vector and targets the victim with the best possible attack (based on the information collected).

\begin{figure}[ht]
\begin{center}
\includegraphics[width=\columnwidth]{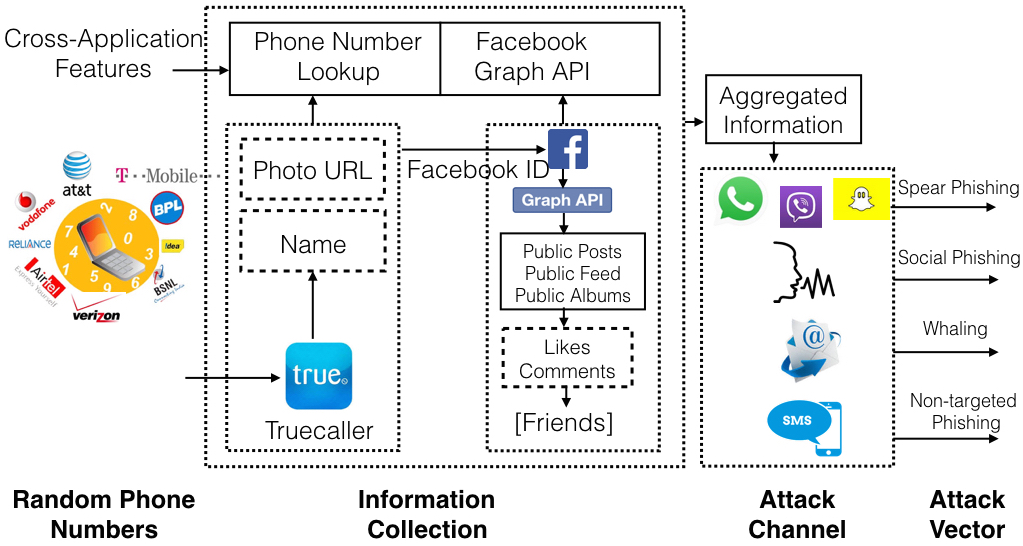}
\caption{System for Cross-Application Information Gathering and Attack Architecture.}
\label{fig:system}
\end{center}
\end{figure}
 
\subsection{Step 1: Setting Up Attack Device} \label{sec:forge-address-book} 
This section elaborates phone number generation and setting up the device under attacker's control, once phone numbers are generated. 
The system generates a large pool of phone numbers which could be 
exploited by an attacker to launch targeted attacks. 
There are several methods to obtain a pool of phone numbers;
consolidating white-pages directory or any other public online directories, 
or scraping the Internet using regex patterns. 
We chose the easiest method for an attacker; 
taking random phone numbers as initial seeds, incrementing 
the digits by one to obtain a sufficient pool. 
Unlike e-mail addresses, 
the phone number set is finite, therefore, an entire range can be enumerated 
and inserted into the address book. This may give a few misses as some phone numbers 
may not be allocated for general use.
Once phone numbers are generated, attacker initializes the address book of the device under his control. The phone numbers added in the address book are now his potential victims for carrying out various kinds of targeted attacks as demonstrated in this paper.

\subsection{Step 2: Collecting Information for an Attack Vector} \label{sec:attack-vector}
In this step, the system aggregates all the available information 
to launch an attack against the victim.
To obtain information about the victim, we used Truecaller, 
an application that enables searching contact information 
using a phone number~\cite{truecaller}. 
Its legitimate use is to identify incoming callers and 
block unwanted calls. 
It is a global collaborative phone directory that keeps data of more than one billion 
people around the globe. 
We used Truecaller as an example, but any such application
can be used to determine this information. Truecaller also maintains data from social 
networking sites and correlates this information to create a large dataset for people who register on it. 
Also, due to its address book syncing 
feature, it retrieves information about contacts (friends) of the ``owner of the phone number'' 
who installed it too. 
The `search'  endpoint of Truecaller application provides details of an individual like: 

{\footnotesize
\begin{quote}
\textit{name, address, phone number, country, Twitter ID, e-mail, 
Facebook ID, Twitter photo URL,} and \textit{photo URL}.
\end{quote}
}

However, the private information obtained is according to the privacy 
settings of users. 

We automated the whole process of fetching information about phone 
numbers from Truecaller. 
We exploited the search end-point (used to search information about a random phone 
number) to obtain the registration ID corresponding to a particular phone number
\footnote{We used this phone number only for research purposes and nothing else.}. 
This was necessary to make authenticated requests and retrieve the information from their servers. 
We extracted the registration ID 
from the network packet sent while searching a random 
phone number on Truecaller application installed on our iPhone as shown in Figure~\ref{fig:nw_packet}. 
Once the registration ID was obtained, we programmatically fetched information for phone numbers in our dataset. 
Multiple instances of the process were 
initiated, on a 2.5 GHz Intel i5 processor, 4GB RAM at the rate of 3000 requests~/~min.
We worked with only one
registration ID for not abusing the Truecaller servers and effecting its 
services, however, it is easy for an attacker to 
scale the process by collecting multiple registration IDs to 
bypass rate limits imposed by Truecaller.

\begin{figure}[htbp]
\begin{center}
\includegraphics[width=.95\columnwidth]{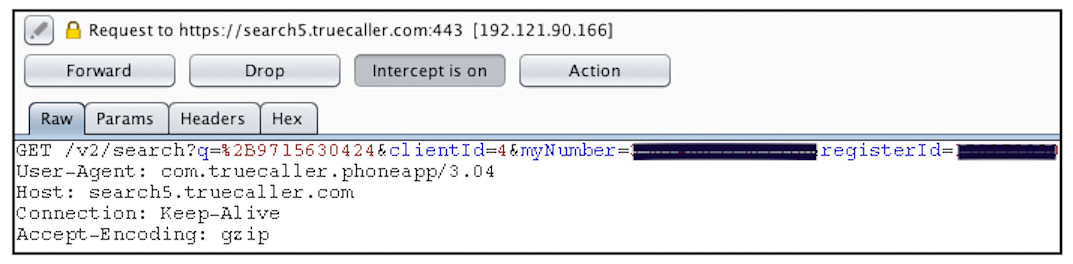}
\caption{Screenshot of a network packet which is used to obtain the registration 
ID from Truecaller to fetch information from its servers.}
\label{fig:nw_packet}
\end{center}
\end{figure}

To obtain the social network of the victim, we used Facebook, the largest social network of family and friends~\cite{facebook-largest}. 
We assume that friends obtained will be related to the person in some way or the 
other which can increase the probability of success of a social phishing attack. 
However, we do not differentiate between the affinity of a 
friend Alice with the victim as compared to another friend Charlie. 
Though, there may be greater affinity with one friend as compared to the other
in the real-world, however, in this paper, we treat 
all friends equal and leave affinity determination as future work. 
Truecaller aggregates data from various social networking websites 
and sometimes provides a link to the public profile picture of the victim on Facebook. 
We extracted Facebook ID from these links 
to retrieve friends of the victim on Facebook. 

Extracting friends from victim's profile is a non-trivial task, 
since everyone does not have their friendlist 
set as public. Therefore, we decided to use public sources 
like victim's public feed, victim's public photo albums, and victim's public posts on 
Facebook to obtain friends information~\cite{graph-api}, 
assuming users liking / commenting on any of these public sources are friends of the victim.
To validate the above hypothesis, 
we performed a small experiment to determine if friends obtained from public 
sources on Facebook are a subset of public friendlist. 
Even though normal 
access token from Facebook does not provide these details, we were able to 
fetch the information using a never-expiring mobile OAuth token obtained from iPhone's Facebook application~\footnote{We are not sure if this is an additional feature provided by Facebook or a bug in their system.
At the time of writing this paper, we did not find any official Facebook documentation about it.}.
We monitored the data packet sent while launching Facebook application on our iPhone 
device and extracted the authentication token to make further requests.

We collected a random sample of 122,696 Facebook IDs and obtained 95,756 friends 
from public sources and 80,979 friends from public friendlist (see Figure~\ref{fig:venn}). 
There were only 62,574 users for whom we were able to find 
friends from both public sources and public friendlist. Out of which, we found that 42,552 (68\%) user-IDs 
liking and commenting on public sources were 
part of victim's friendlist with more than 95\% matching rate. As observed in Figure~\ref{fig:venn}, in some cases friends from public sources were not a complete subset of 
friends from public friendlist. We obtained 5,881 friends with 90 - 95\% matching, 3,754 friends with 85 - 90\% matching, and 10,387 friends with less than 85\% matching.
This could be because some users might have disabled all platform 
applications from accessing their data. In this case, they might not appear 
anywhere in any Facebook API~\cite{fb-friends}.
To launch attacks using friends information, friends can be picked from public friendlist, 
if available, else, the attacker can rely on public sources to extract friends.
Therefore, we extracted the Facebook ID from the photo URL (using regular expression) obtained from Truecaller JSON response,  
and obtained public sources using Facebook Graph API to find friends on Facebook to craft social phishing attack vector. 
For example, following JSON object was obtained for one of the phone numbers in the dataset -- 

{\footnotesize
\begin{lstlisting}[language=json]
{
"NAME": "XXXXX",
"NUMBER": "+91XX0000000X", 
"COUNTRY": "India", 
"PHOTO_URL": "http://graph.facebook.com/XXXXXX/picture?width=320&height=320", 
"e-mail": " "
}.
\end{lstlisting}
}

The Facebook ID was parsed from \texttt{PHOTO\_URL} and used to make further requests.
E-mail addresses for some users were also available which can be used to target them.

\begin{figure}[ht]
\begin{center}
\includegraphics[width=.95\columnwidth]{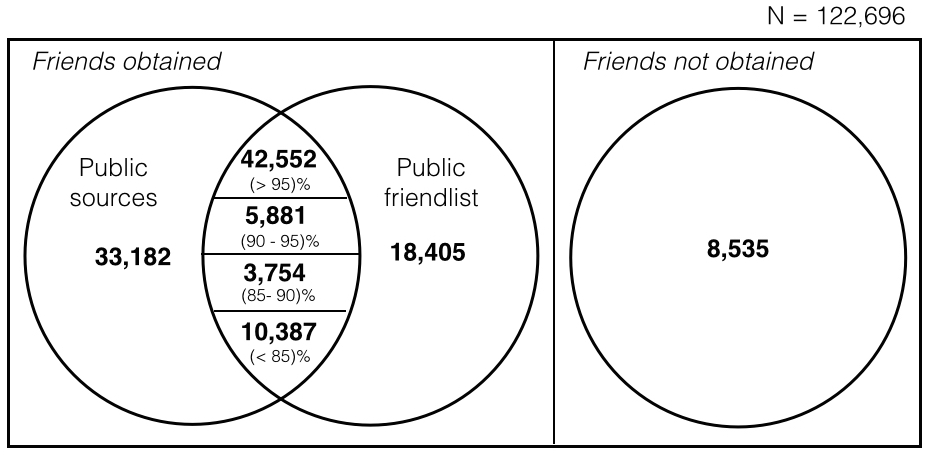}
\caption{Relation between friends obtained from public sources and public friendlist. Friends from public sources are found to be a subset of friends from public friendlist in 68\% cases (with more than 95\% matching).}
\label{fig:venn}
\end{center}
\end{figure}


An attacker can utilize the information, as obtained in this step and craft attack 
vectors described in step 4.
The 
attacker can craft non-targeted attacks, in case no information about the victim is obtained.

Apart from applications like Truecaller and Facebook that we explored in this paper, attackers can exploit CNAM (Caller ID Name) database~\footnote{http://www.voip-info.org/wiki/view/CNAM}, a database that is linked to names of calling number. The service operational in US, provides information associated with a landline number. Attackers can use this to obtain basic information about their targets.
This is out of scope of current work.

\subsection{Step 3: Formulating Attacker's Profile and Attack Channel} \label{sec:attack-channel}
Once the data is collected about phone number's owner, the system determines the channel (\ott, voice, e-mail, or SMS) to phish the victim. This entirely depends on
whether the victim is present on that particular channel.
\paragraph{\ott}  
If the attacker decides to choose \ott like WhatsApp, Viber, or Snapchat, he needs to ensure if the victim is using one of these applications. 
This is achieved by exploiting the address book syncing feature in \ott. Once a user registers himself on these applications, 
his contacts in the address book 
are uploaded (automatically, for some applications) to the \ott' service provider and 
are matched against the users of the application to find already existing contacts. Only 
the information about the owners of the phone numbers present in the address book is 
retrieved. Unlike Facebook and Twitter, these applications make no suggestions / 
recommendations for people who might be using these \ott.
While this makes it easy and convenient for users to discover friends on these applications 
rather than adding them manually, it poses a security threat as well as, an attacker can use 
this to find the presence or absence of the victim on these applications (i.e., the attack channel).
The attacker can himself create a convincing profile (profile picture and local phone number) on WhatsApp to make sure the victim feels that attacker is legitimate.

\paragraph{Voice} 
In addition, an attacker can choose voice as an attack channel to target their victims.
Similar to \ott, to target victims on this channel, an attacker needs to gather relevant information without checking the presence of the victim beforehand. 
An attacker can device a strategy to make himself look like a trusted or legitimate organization, to gain victim's trust. 
Specifically, attackers can exploit the integrity of information provided by \caller like Truecaller~\cite{truecaller}, Facebook's Hello~\cite{hello}, and Whitepages Caller ID and Block~\cite{current_caller_id}. These applications are emerging to help in getting additional information about the incoming caller. 
In general, these applications allow an individual to register 
using his~/~her phone number and help in identifying 
the caller by showing the information (like name) from their respective databases.
\Caller also gather information from social networking sites to collect 
more information about the caller. 
Attackers can undermine such an application to gain trust of their victims. 
Specifically, a) attackers can register a phone number (controlled by him / her) 
as a trusted bank / company / organization in which a user is interested in or is dealing 
with; b) spoof one of the already registered phone numbers with the 
\caller and call victims such that the call appears to come from a real entity. 
We describe each of them in detail.

\begin{itemize}
\bitem{Fake Registration}
An attacker can add spurious information in \caller fairly easily, 
thus compromising the integrity of the information provided by them to gain victim's trust. 
Associating an identity with a phone number increases the trust of an individual and likelihood to pick a call. 
Since \caller do not have a mechanism 
for verification of the users' details, and rely on the information provided 
by the user while registering, it is easy for an attacker to abuse this trust. 
For example, an attacker can register as multiple fake banks on \caller as shown in Figure~\ref{fig:tc_hdfc}. We only use HDFC Bank as
an example, but any name / entity can be used.

\begin{figure}[hb]
\begin{center}
\includegraphics[width=3.21cm, trim=4.9cm 6.8cm 3.6cm 0cm, clip]{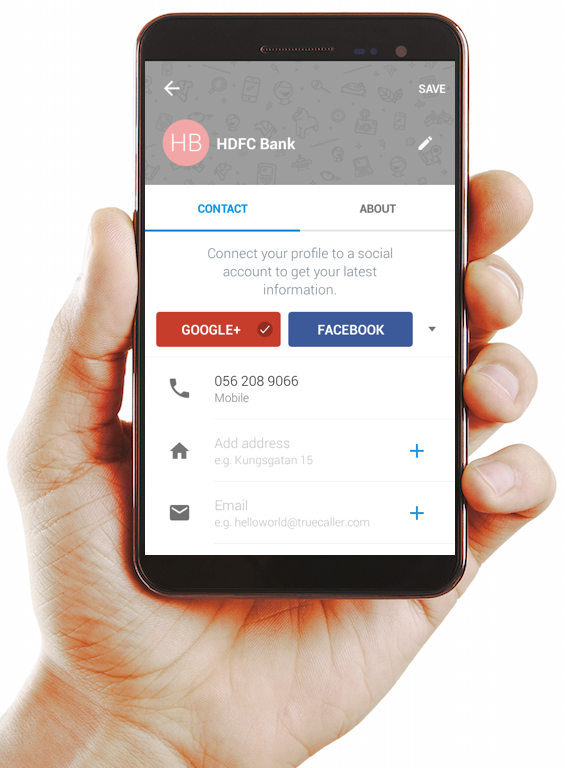}
\caption{Fake registration on Truecaller as HDFC.}
\label{fig:tc_hdfc}
\end{center}
\end{figure}

For registration, he needs a smartphone device with working phone connection. 
It is a manual process where a short SMS code is sent from \caller to verify the phone number.
Since the number of banks are limited, it is not difficult for the attacker to do this manually.
Attacker does not know the bank of all these victims, therefore, 
the attacker can target a large user population to achieve a good success rate. 
As millions of automated VoIP calls can be generated at low cost, 
as shown in the past, attacker does not need to worry about picking only the victims whose banks are known~\cite{griffin2008vishing}. 
To make the profile look more authentic, 
fake social media profiles can be created 
and linked to \caller while registering an account on it. 
Top five most popular
applications (Truecaller, Whitepages caller ID and Block, Facebook's Hello, Whoscall-Caller ID
and Block, and Contactive) with fake registrations as HDFC bank by exploiting
\caller has been shown in Appendix. 

\bitem{Caller ID spoofing}
Another trick to deceive victims uses caller ID spoofing which can be carried out by imitating already registered 
phone numbers or other phone numbers
whose details were uploaded by \caller exploiting address book syncing feature. 
As a user must have entered some details about 
him / her while registering, it makes the attacker look genuine, 
than an unknown phone number flashing on the screen.

\end{itemize}

\paragraph{SMS} 
An attacker can choose SMS as the attack channel to phish their victims. He needs to gather relevant information about the victim without checking his / her presence on the attack channel. 
As the victim does not see any profile of the attacker, he can choose using a local phone number to gain victim's confidence.

\paragraph{E-mail} 
Since so many people around the world depend on e-mail, it is a lucrative channel for attacks. Attackers lure people in giving away their information or entice them to take some action. If there is a non-empty field in the JSON object received from the Truecaller, e-mail can be used to phish users. Attackers can craft these e-mails to look convincing, sending them out to literally millions of people
around the world~\cite{email-million}. 
As attacker's profile is visible to the victim, he can carefully choose an e-mail address and name to convince victim about his authenticity.


\subsection{Step 4: Crafting Attack Vector} \label{sec:attack-vector}
After the attack channel is determined, attacker can craft appropriate 
attack vector to phish the victim.
We describe the attack vector generation details for each of the attacks below.

\subsubsection{Social Phishing Attacks}
Although phishing is a social engineering attack, 
here we discuss social phishing~\cite{jagatic2007social}, i.e., how
phishing attacks can be better targeted by making them appear 
to be coming from a friend within victim's own network. 
Friends' information can be conveniently chosen to gain trust, therefore, 
the attacker uses victim's name and one of his friend's 
information (i.e., friend's name) to craft the attack vector. 
This information is obtained from Facebook, as discussed earlier in this section.

\subsubsection{Whaling attacks}
Whaling attacks~\cite{whaling} that are directed specifically at senior executives or other high-profile 
individuals within a business, government, or other organization. It uses 
the same technique as above mentioned targeted phishing, vishing, or whaling attacks but the intended victims are people with high influence
or high-net-worth individuals.
In India, there is a particular set of phone numbers reserved by mobile operators for politicians, 
bureaucrats, and people willing to invest large amount of money to get these special phone numbers. 
They are called Vanity / VIP / Fancy numbers and follow a specific pattern~\cite{vanity-pattern}. 
It could be one digit repeated several times, \texttt{99999-xxxxx} or \texttt{xx-8888-xxxx}; 
two digits, \texttt{xx-85-85-85-xx}; or in different orders, \texttt{xx-123-123-xx} or \texttt{xx-11-112233}. 
The main advantage of vanity phone numbers over standard phone numbers is 
increased memorability. Since they 
are bought at higher price, owners of these phone numbers can be assumed as people with high influence who likely are more attractive targets for attackers~\cite{vanity}. 
For very special numbers, network providers 
host auctions online where people can purchase these numbers~\cite{bsnl-auction}. 
Using only vanity numbers in the address book, attackers can launch 
whaling attacks that only targets 
HNIs (High-net-worth individuals) by sending them 
targeted or non-targeted phishing messages or initiating vishing calls.

\subsubsection{Spear Phishing Attacks}
Spear phishing attacks are directed at specific individuals or companies. 
These attacks are crafted 
using some a-priori knowledge of either victim's 
name, location, or interests to make it more believable and increase the likelihood of its success.
We focus on generating spear phishing 
attack vectors using victim's name, as obtained from Truecaller. 

\subsubsection{Non-targeted Attacks}
Non-targeted attacks are undirected attacks which are aimed to target as many users 
as possible. The goal is to reach out to a large audience and not to target 
a particular individual. Since it only requires the knowledge whether the victim 
is present on the channel, this can be achieved 
by crafting a non-targeted phishing, vishing, or whaling attack, even if no information about the victim is available.
With low VoIP calls, non-targeted vishing attacks are cost-effective for attackers.

All the targeted and non-targeted attacks described in this step can be launched on either channels viz., \ott, voice, e-mail, or SMS.

\section{Scalability} \label{sec:scalability}
To define \textit{scalability}, we assume that an attacker starts with no information about its potential targets.
The attack method's scalability can be characterized  by the fraction of people who can be reached over an attack channel, and targeted attacks can be launched against them.
To demonstrate the scalability of our attacks, we enumerated through a 
list of 1,162,696 random Indian phone numbers. Since these numbers are chosen randomly, no
additional information is available about their owners at the beginniing.
We demonstrate the scale at which each of the proposed attacks can be carried out with the techniques
described earlier.

\subsection{Phishing Attacks} 
We forged the address book of an Android device by inserting all  
these phone numbers in multiple phases. 
The next step was to collect attributes associated with the owner of the phone number (victim). 
Truecaller (TC) was used to collect more information about the victims. 
Detailed information for 722,696 (62\%) users was collected using Truecaller; name was obtained for all the users as shown in Figure~\ref{fig:scale}.
For rest of the users whose information cannot be 
obtained from Truecaller, non-targeted phishing attacks can be launched against them.
To craft more targeted and personalized attacks, i.e., social phishing attacks, 
friends information was leveraged from Facebook (FB). 
Social circle information was obtained for 114,161 (93\%) out of 122,696 users; 80,979 from public friendlist and 33,182 friends from public sources. 
To check the presence of these numbers 
on an attack channel, they were synced with WhatsApp application (WA) using 
address book syncing feature. About 51,409 users were present on WA.
Social phishing attacks can be launched against these users whereas spear phishing attacks 
can be launched against other 180,000 users whose social circle was not obtained, but were present on WA. Numbers which were not found on WhatsApp either may not be allocated to any user or may not be registered on it. 

Spear phishing attacks can be launched against 600,000 users on voice or SMS. In addition, 122,696 users can be social phished on voice or SMS.
E-mail address for 81,389 users were obtained from Truecaller; 13,754 can be social phished and 67,635 can be spear phished on voice or SMS.
\begin{figure}[ht]
\begin{center}
\includegraphics[width=0.48\textwidth]{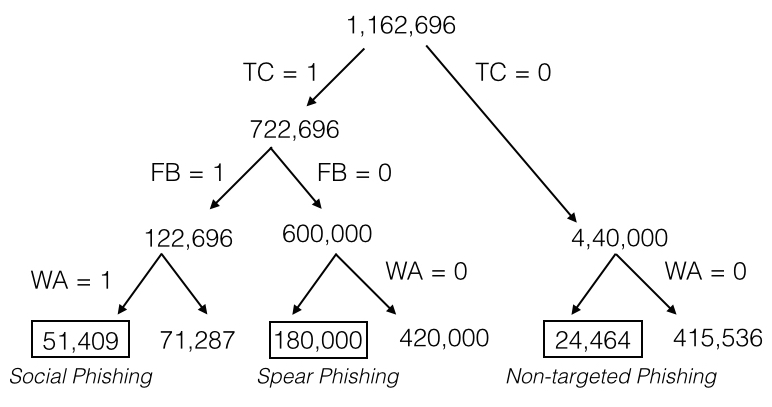}
\caption{Data collection to demonstrate scalability of phishing attacks of the system choosing \ott as the attack channel, WA--WhatsApp, TC--Truecaller, FB--Facebook.}
\label{fig:scale}
\end{center}
\end{figure}

\subsection{Vishing Attacks}
Personal information for 722,696 users was found on True- caller against 1,162,696 phone numbers searched. Vishing attacks can be crafted against the owners of these phone numbers. We could extract Facebook ID's for 122,696 users. Using Facebook Graph API, we obtained following details for these users: gender (112,880), relationship status (57,755), work details (92,352), school information (110,426), employer details (106,746), birthday (9,728), and hometown (80,979). The collated information can be used to increase the success rate of targeted vishing attacks.

\subsection{Whaling Attacks}
As owners of vanity numbers might belong to elite
members of the society, they can be of particular interest to
attackers. 
We looped through the ``patterns" available 
from an e-auction website to enumerate vanity numbers pool~\cite{bsnl-auction}. 
We initialized our smartphone's address book with 171,323 vanity numbers. 
We found 91,487 vanity numbers on Truecaller and 11,286 on Facebook. 
They were synced with WhatsApp and 5,756 (51\%) were found 
on it.
Out of 11,286 vanity numbers that were found on Truecaller and Facebook; we obtained personal information (using Facebook) about owners as follows: gender~(10,246), relationship status~(3,733), birthday~(726), work details~(6,729), school details~(10,994), employer details (9,801), and hometown~(6,952). 
We manually analyzed Facebook profiles of 100 random vanity number owners to find their occupation details and found director~/~CEO~/~chairman~(10), student~(10), engineer~(12), consultants~(2), business~(5), accountant~/~officer~(8), lecturer~(5), manager~(8), bank officials~(12) for 70 user profiles.

Whaling attacks with social information was obtained for 11,286 users which can be attacked on voice or SMS. However, only name was obtained for 80,201 users, using Truecaller, who can be made targets on voice or SMS.
E-mail address for 11,013 users was obtained; 1,354 users with social information. Attacks can be made more targeted and personalized against these users.


\section{Ethical and Legal Considerations} \label{sec:ethical-legal-consideration}
Crawling data is an ethically sensitive area. We did the data collection just to 
demonstrate the feasibility and scalability of targeted attacks by abusing a phone number. 
The goal of this work was not to collect personal information 
about individuals, but to explore how such applications can be abused to collect personal information. 
As a proof of concept, we collected information about 
owners of random phone numbers ensuring that the collected 
information is not made available to any other organization or individual. 
All the conducted experiments were approved by the Institutional Review
Board (IRB) of our institution. Data collected from the participants was
anonymized and protected according to the procedures described
in the corresponding IRB submission documents.
At the end of our experiments and analysis, phone numbers of all the profiles were delinked to maintain privacy and confidentiality of data. Phone numbers and Facebook IDs were hashed to preserve privacy.
We collected only the public information available on Facebook using it's Graph API.
\section{Study Design} \label{sec:study}
In this paper, we only focus on demonstrating the effectiveness of phishing attacks on 
\ott. Thus,  we do not address effectiveness of the other attacks and channels as 
demonstrated in previous sections.

To demonstrate the \textit{effectiveness} of phishing attacks on \ott,
we chose to conduct an online roleplay user study on Amazon Mechanical Turk (MTurk). We chose 
this over a real-world study because conducting a real-world study would involve user deception. 
To launch attacks as proposed in this paper would require us to gather personally 
identifiable information (phone number) that would not be possible without deception. 
Deception studies challenge the principle of \emph{respect for person}, and are generally 
not preferred in the research community due to unforeseeable consequences~\cite{department2014belmont, jakobsson2006designing}. 
Debriefing, unless carried out face-to-face between researcher and the subject, 
remains a challenge as it is almost impossible to replicate it in online phishing studies.
A study that launched actual phishing attacks without informing their users reported 
that users got infuriated with the attacks after they were debriefed~\cite{jagatic2007social}. 
Owing to these constraints and limitations, we chose to validate the effectiveness of our attacks using a roleplay user study.
The benefit of the roleplay 
is that it enables researchers to study effect of phishing attacks 
without conducting an actual phishing attack. 
It has been shown that such roleplay tasks have good internal and 
external validity~\cite{downs2006decision, kumaraguru2010teaching, sheng:who-falls-for-phish:-a-de:2010:lrfkq}. 
The study was built using Javascript framework and not based on 
standard self-reporting data collection techniques like survey and questionnaires.

Participation was restricted to those who were above 18 years of age and 
had been using WhatsApp on regular basis. 
There was no restriction on the participants belonging to a particular region or country. 
All participants who completed the experiment were paid \$0.30. 
Our experiment consists of 
two phases, 
a) Briefing: to ensure that participants 
have concrete information and clear role description. 
b) The Play: to assess participants susceptibility
to phishing attacks. 
Participants were exposed to one of the three phishing attack vectors: 
non-targeted (\eone), spear (\etwo), and social (\ethree).
%
%


\subsection{Briefing} \label{sec:role-assignment}

Susceptibility to phishing attacks was 
measured with response to a roleplay task which was built on Javascript and showed multiple screens as WhatsApp screenshots. 
This phase of the study was common across all participants to 
familiarize them with the real-world scenario. 
A hypothetical situation for the participant to assume himself as ``Dave'', 
and has friends ``Alice'', ``Bob'' and ``Charlie''. 
We used Facebook as a medium to bootstrap this
and to let the participant familiarize with the roles. 
An attacker can 
extract this information from Facebook and create a social phishing
attack vector to phish the victim, which is modeled in this step.
Next screen shows the registration on WhatsApp, as ``Dave'', 
using a random phone number.
The aim is to make the participant 
understand that the user (i.e., Dave) has an account on WhatsApp.
Once registered, WhatsApp syncs Dave's address book and found only Charlie. 
Note that other Facebook friends of Dave (i.e., Alice and Bob) were not present on WhatsApp.
There could be two reasons, either Dave does not have Alice's and Bob's number
in his smartphone or Alice and Bob do not have an account on WhatsApp. 
The primary idea is to introduce 
Charlie as the \textbf{only} friend who is present on WhatsApp and other friends 
(Alice and Bob) are not present on WhatsApp. This is to mimic
the attack model when an attacker might not  know which of victim's friend
is present on WhatsApp. Therefore, the success of the attack should be independent of 
this knowledge; who provided wrong answers to any of the following two questions 
asked, 
``\textit{Who was your friend on Facebook?"}, 
and ``\textit{Who was your friend on WhatsApp?"}. Since correct answers to these questions were provided during course of the experiment, participants who did not provide correct answers were filtered out.

\subsection{The Play} \label{sec:roleplay}
In this phase of the user study, participants were 
exposed to one of the three cases of phishing: non-targeted, spear, and social. 
Each of the case 
was randomly assigned to the participants. 
In all three cases, the 
legitimate case was shown to the participant
to ensure that his responses were as expected as 
in a real-world scenario i.e., given a message \emph{m}
and a trust function \trust,
\begin{equation}
\small
\label{eq:trust}
\trust(m)\ from\ known\ no. \geq
\trust(m)\ from\ an\ unknown\ no.
\end{equation}

The order of phishing and legitimate messages was randomized to avoid learning bias during the course of the 
experiment. 
Some participants were removed from the analysis due to 
unexpected behavior as discussed in Section~\ref{sec:results}.
At the end of each WhatsApp message shown to the participant, 
the participant was asked the following question and corresponding options:
\emph{``What would you like to do with the message?"} with following options: click, reply, delete, or do nothing.
Now we describe the three experiments to test the success of phishing attacks. Since the names and message content were kept same in all the three experiments, we do not foresee any bias. 

\paragraph{\eone: Testing non-targeted phishing attack's success}
Non-targeted phishing is defined as an attack scenario where no additional information 
about the victim is known beforehand, except the phone number. 
In the \textbf{play phase}, participants 
were exposed to two scenarios in a random order to avoid learning bias; 
\textit{probably phishing message} (\prob[1]) and \textit{legitimate message} (\legit[1]) . 
In \prob[1], the sender is a random phone number, whereas, in \legit[1], 
the message is from Dave's friend Charlie. 
The former scenario is \emph{probable phishing} because from Dave's perspective, the sender 
could be one of his friends who is not present in his WhatsApp contacts. However, the latter case is 
\emph{legitimate} because Charlie was already in Dave's address book, 
as mentioned during the briefing phase (see Section~\ref{sec:role-assignment}). 

\paragraph{\etwo: Testing spear phishing attack's success}
Spear phishing is defined as an attack where some information about the victim is known 
beforehand, in addition to his phone number. In our experiment, name of the participant was 
``Dave", as described in the briefing phase. This information was used to craft a spear phishing 
attack vector. Similar to non-targeted phishing, participants were exposed to two scenarios; 
\textit{probably phishing message} (\prob[2]) and \textit{legitimate message} (\legit[2]).
In \prob[2], the sender is a random phone number, whereas in \legit[2], the message 
appears to be coming from the friend Charlie. However, with one notable difference, 
that the name of the victim (Dave) was added to 
the message to make it more personalized as compared to \eone.

\paragraph{\ethree: Testing social phishing attack's success}
Social phishing is defined as an attack where social information 
(friends, acquaintances, colleagues, etc.) associated with the victim is gathered, 
in addition to known basic information about the victim (name and phone number). 
In this part of the experiment, participants were exposed to three scenarios 
(as compared to two in \eone and \etwo) in a random order; 
\textit{probably phishing message} (\prob[3]), \textit{legitimate message} (\legit[3]), and \textit{phishing message} (\phish[3]). 
In \prob[3], the sender is a random phone number, however, mentioning 
the name of ``Alice'' (one of Dave's friend on Facebook but not 
in the WhatsApp contacts, see Section~\ref{sec:role-assignment}). 
From Dave's perspective, this could probably be a legitimate message 
because Alice is not in Dave's address book and plausibly in real-world scenario Alice is 
trying to initiate a conversation with Dave. On the other hand, 
this could be a phishing message, because
friend's name (Alice) could be forged and an attacker could 
imitate Alice and send a message to Dave.
In \legit[3], the message appears to be coming from the friend Charlie and in \phish[3], the message 
is coming from a random phone number having Charlie as the friend's name. 
Since, Charlie is anyways a friend of Dave on WhatsApp, 
this is definitely a phishing attack because the sender phone 
number should have shown Charlie and not a random number. 

\begin{figure*}[ht]
\centering
\subfigure[{\prob[1]}]
{ 
 \includegraphics[width=4cm, frame, trim=5.1cm 17cm 4cm 1.8cm, clip]{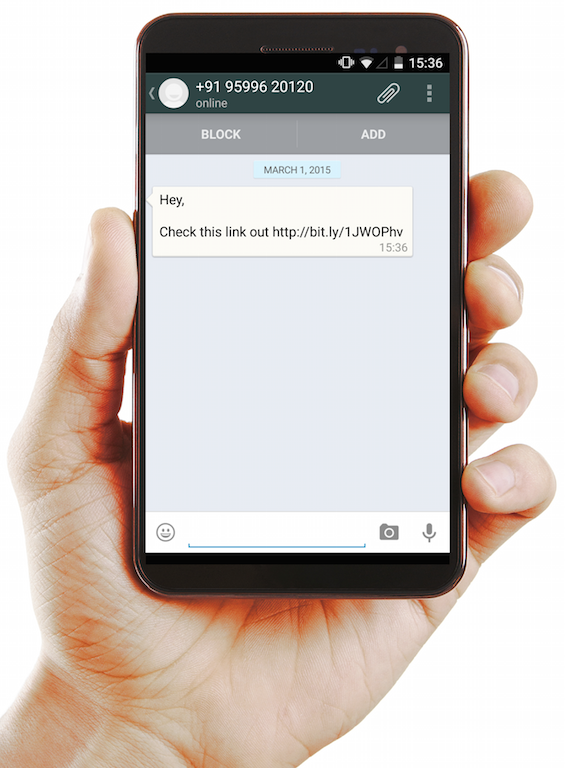}
\label{fig:random-vul}
}
\subfigure[{\legit[1]}]
{ \includegraphics[width=4cm, frame, trim=5.15cm 17cm 4cm 1.8cm, clip]{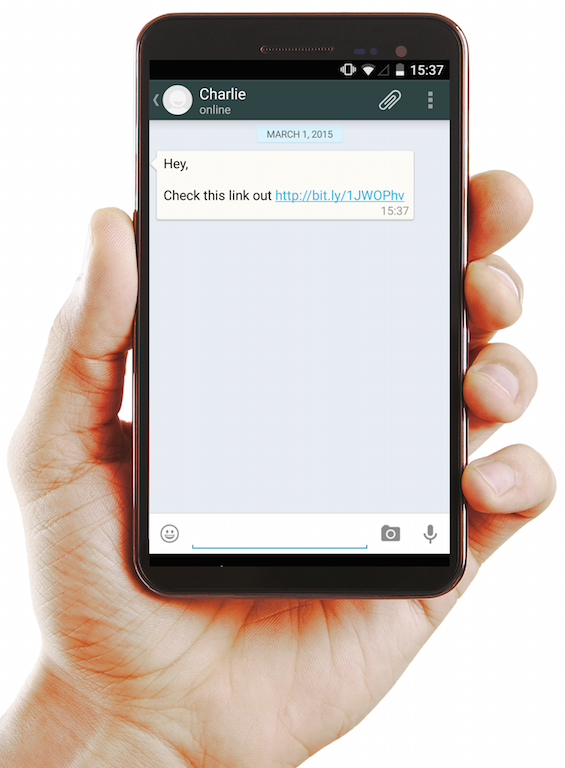}
\label{fig:random-leg}
}
\subfigure[{\prob[2]}]
{ \includegraphics[width=4cm, frame, trim=5.1cm 17cm 4cm 1.8cm, clip]{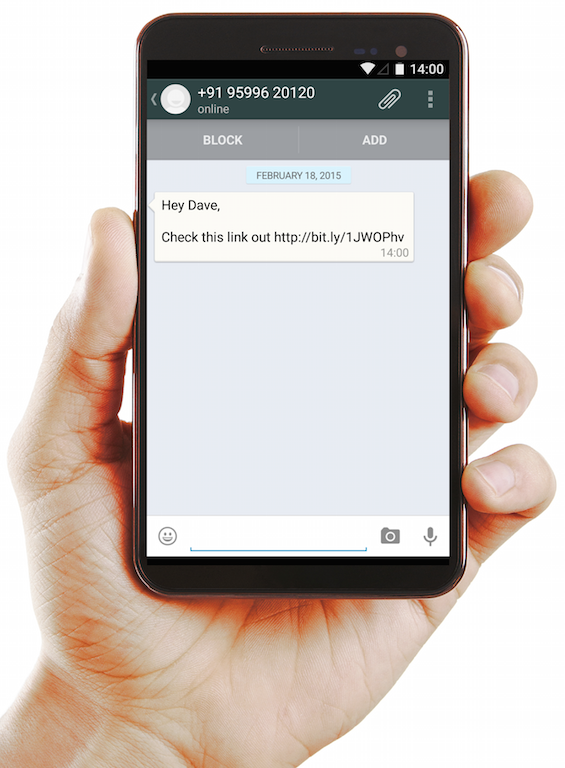}
\label{fig:spear-vul}
}
\subfigure[{\prob[3]}]
{ \includegraphics[width=4cm, frame, trim=5.1cm 17cm 4cm 1.8cm, clip]{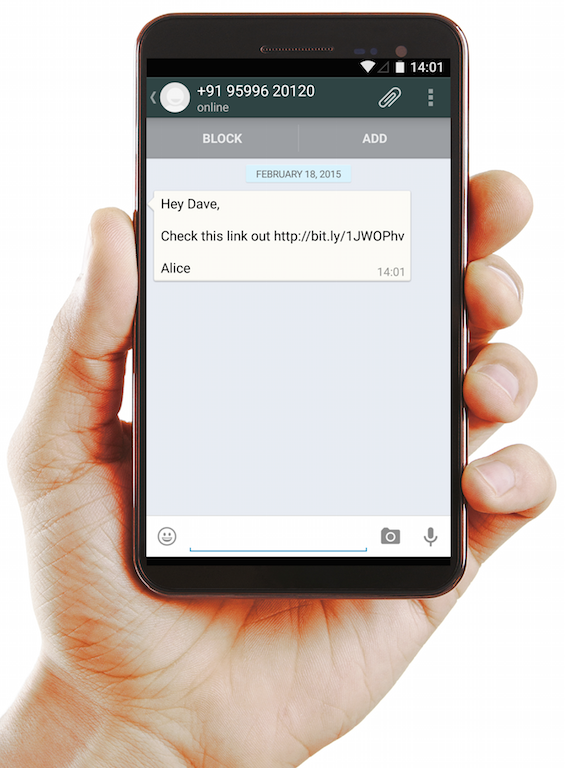}
\label{fig:social-vul}
}
\caption{Three phishing attack scenarios shown to the participants (\textit{here, we only show the trimmed version of the actual screenshots}). The top left-hand corner shows the sender's phone number (or name in case the sender's phone number is in WhatsApp contacts).}
\label{fig:social}
\end{figure*}

\subsection{Effectiveness of Attacks} \label{sec:results}
In this section, we demonstrate the \emph{effectiveness} 
of phishing attacks performed during our user study. 
In total, 460 participants completed the entire user study, 
out of which 129 participants were filtered out based 
on answers to two questions asked from the participants during the briefing phase
(see Section~\ref{sec:role-assignment}). We present the results based 
on remaining 331 participants. 
We used Kruskal Wallis and Mann Whitney statistical tests to 
check the behavior of population subject to different order of messages shown to them. 
We did not find statistical difference between the random
groups in falling victims to all three kinds of phishing attacks. 

Table~\ref{tab:all-cases} summarizes the results obtained 
in three experiment scenarios \eone, \etwo and \ethree
as mentioned in Section~\ref{sec:roleplay}. 
In this work, we define the success of a phishing attack
if either a participant decided to click 
the link or reply to the message, whereas, 
attack is unsuccessful if the participant decided to delete or do nothing about 
the message. \emph{Note that these are potential victims 
who may fall for phishing attacks.
Actual phishing attack happens when the participant
goes through all the steps in the phishing attack i.e. by providing his / her personal details like credit card information, passwords etc.
on the phishing web page.}
However, previous studies have established that a very high percentage of participants 
who click on the link continue to provide information to the phishing 
websites~\cite{kumaraguru2009school, kumaraguru2010teaching, sheng:who-falls-for-phish:-a-de:2010:lrfkq}. 
We believe that users who choose to reply to the message are \textit{potential victims} too, 
as the attacker can verify active usage of the phone number. Also, attacker can 
lure the victim to give out personal information in subsequent messages. 
Extra cautious users would have preferred to either delete / do nothing with the message received.
We denote,

$\probclick, \legitclick, \phishclick \implies$  clicked / replied to the message, and

$\notprob, \notlegit, \notphish \implies$  deleted / did nothing about the message

For example, \probclick means participant chose to click or reply to a \textit{probably phishing message}, 
while \notprob means participant chose to delete or do nothing about the \textit{probably phishing message}. 
We remove those participants from our further analysis who chose to 
click on phishing / vulnerable message
but not on the legitimate message. Because according to equation~\ref{eq:trust},
$
\trust({\legit}) \geq \trust({\phish})
\quad and \quad \trust({\legit}) \geq \trust({\prob})
$.
We denote these participants
as \emph{Unknown} (see Table~\ref{tab:all-cases}).

\begin{table}[ht]
\centering
\setlength\extrarowheight{2pt}
\caption{Possible outcomes of phishing attacks for each experiment. (\prob $\rightarrow$ probably phishing message, \legit $\rightarrow$ legitimate message, \phish $\rightarrow$ phishing message, \legitclick $\rightarrow$ click / reply to a legitimate message, \notlegit $\rightarrow$ delete / do nothing to a legitimate message).} 
\begin{tabular}{| c | l | l | l |} \hline
\textbf{Case} & \textbf{Vulnerable}  & \textbf{Cautious} & \textbf{Unknown}  \\ \hline\hline
\multirow{2}{*}{\eone} & \probclick\legitclick (37) &  \notprob\legitclick (24) & \probclick\notlegit (7) \\
& &  \notprob\notlegit (46) &     \\\hline
\multirow{2}{*}{\etwo} & \probclick\legitclick (56) & \notprob\legitclick (19) & \probclick\notlegit (4)     \\
& &  \notprob\notlegit (28) &      \\\hline
\multirow{3}{*}{\ethree} & \probclick\legitclick\phishclick (54) & \notprob\legitclick\notphish (12)  & \probclick\notlegit\notphish (2)    \\
& \probclick\legitclick\notphish (9) & \notprob\notlegit\notphish (20) & \notprob\notlegit\phishclick (1)   \\
& \notprob\legitclick\phishclick (9) &  & \probclick\notlegit\phishclick (3) \\ \hline
\end{tabular}
\label{tab:all-cases}
\end{table}

We denote those participants who chose to click / reply on either phishing (\phish) or probably phishing messages (\prob)
or both as \emph{Vulnerable} 
(i.e. falling for phishing attacks). 
All other participants were part of \textit{Cautious} group, i.e., who 
chose to delete or do nothing about both  phishing (\phish) and probably phishing messages (\prob).

We define the success
rate of phishing attack as:

$$Success (\%) = \frac{Vulnerable}{Vulnerable + Cautious} * 100$$

In total, we have 314 out of 331 participants 
who were either vulnerable or cautious. 
We rest our analysis on these 314 participants. 
We found that phishing attacks on \ott were 
successful as, \eone~=~34.5\% (37 out 107), \etwo~=~54.3\% (56 out of 103), and \ethree~=~69.2\% (72 out of 104). 
This is \emph{consistent} with prior work that social phishing is the most 
effective out of the three. Furthermore, in social phishing
as observed from Table~\ref{tab:all-cases}, 
equal number of participants (63 = 54+9) fell for phishing 
when the name mentioned in the message text was Charlie (i.e., it is coming from a friend who is 
in Dave's WhatsApp contacts) and when the name was Alice 
(i.e., it is coming from a friend who is not in Dave's WhatsApp contacts). 
This shows that including friend's name in the message (irrespective of whether the 
friend is present or absent on WhatsApp) increases the success rate of phishing attacks.
We repeated the analysis for random 25\%, 50\%, and 75\% of the total participant population and found the success results to be consistent. Hence, we can establish that the participant pool size is sufficient for our analysis.

\subsection{Limitations} \label{sec:limitation}
There are a few limitations to the current study. 
First, the sample was drawn from MTurk users and is not expected to 
be a complete representative of people using \ott. Our sample of MTurk 
users tend to be younger, more educated, and more tech-savvy than the general public.
A second limitation of this study is the lack of direct consequences for user 
behavior. Participants might be more willing to engage in risky behavior in 
this roleplay if they feel immune to any negative outcomes that may follow. 
However, the factors used to determine phishing susceptibility 
would not differ as observed in the real-world behavior.

\section{Mitigating Risks} \label{sec:recommendations}
Based on the exploits we found as described in Section \ref{sec:attack-vector}, we present recommendations on how to alleviate (if not eliminate) the security risks created by these exploits. 
We communicated with a number of service providers highlighting the security issues; they considered it to be a serious problem and ensured prompt redressal. 
To combat the abuse, there have been services in place, for instance, CNAM lookup databases~\footnote{http://www.voip-info.org/wiki/view/CNAM} for landlines numbers.
Also, initiatives like Secure Telephone Identity Revisited (STIR) working group~\footnote{https://datatracker.ietf.org/wg/stir/charter/} aim to provide a more secure telephony identity by limiting the ability to spoof phone Caller ID. However, this relied on PKI that is currently not implemented.
Recently, some services have put some defensive measures in place. 
WhatsApp incorporated spam blocker feature 
as a first step in this direction~\cite{whastapp-spam}, though their effectiveness need to be studied.
Facebook patched the mobile OAuth token which enabled us to obtain personal public information, which was otherwise unavailable via Graph API.
\subsection{OTT Messaging Applications}
Given the plausibility of phishing attacks on \ott, this medium needs to be better defended. 
\ott can put certain checks; restrict address book sync features, where people can be added only based on requests (like Facebook). 
End-to-end encryption poses a major challenge to identify a phishing message at zero hour. In order to effectively defend against phishing message, one solution could be assigning crowd-sourced score (phishing) to the source of a phishing message. 
\ott can filter messages with high phishing score. 
However, introducing noise in the dataset, remains a challenge. 
To this end, we are currently investigating a defense from our side. It involves building cross platform intelligence from existing Internet infrastructure to associate history with a given phone number which can hep in modeling potentially bad phone numbers.

\subsection{Caller ID Applications}
There is also a necessity to ensure the integrity of the information provided by \caller,  
as people rely heavily on them to know about the incoming call and trust the information provided by these services. 
\paragraph{Verification}
One of the biggest challenge that \caller have to face is to implement verification of the information provided by users. 
Currently, at the time of registration, only phone numbers are verified, 
and neither the entity behind these phone numbers nor the details of 
owners of these phone numbers is verified.
 \Caller can check the integrity of specific business organizations with appropriate authority, listing them as verified users, and routinely scan for any malicious activity in these accounts. 
Similar techniques can be implemented by \caller to maintain 
the integrity of the information stored in their databases.

\paragraph{Additional Information About Callers}
Additional information can be provided about the caller / the owner of the phone number. 
For instance, applications can record 
the timestamp when the account was registered and call frequency patterns. 
These details can be provided to the user 
so that he / she can make an informed decision about the caller. 
In addition, 
social information about the caller can be displayed, like number of 
mutual friends, presence on social networks etc.
Caller ID applications can design several metrics 
like social rank based on the information 
aggregated across social networks. If the same name 
appears across multiple networks and the user is 
found to be active, he / she can be assigned a higher score than a passive user.

\paragraph{Delinking information}
Caller ID applications like Truecaller serve as a reservoir of information, by collating information from multiple sources. 
The data from different sources should not be aggregated and stored at the same place, it serves as a goldmine reservoir. Some parts of the information can be even encrypted to ensure unnecessary information leakage.
\section{Conclusion} \label{sec:conc}
OTT messaging and Voice over IP applications are gaining popularity worldwide. These applications have millions of registered users. 
As much as these applications attract users, spammers find them attractive as well. 
In this paper, we demonstrated the \textit{feasibility, automation, scalability} of targeted phishing, vishing, and whaling attacks by abusing phone numbers. 
We investigated how easy it would be for a potential attacker to launch automated targeted and non-targeted attacks on different channels viz., \ott, voice, e-mail, or SMS. 
We presented a novel, scalable system which takes a phone number as an input,
 leverages information from Truecaller (to obtain vicitm's details) and Facebook (to obtain social circle), checks for the presence of phone number's owner on the attack channel (\ott, voice, e-mail, or SMS), 
 and finally targets the victim. 
We collected information for 1,162,696 Indian phone numbers and 
show how non-targeted and targeted phishing, vishing, and whaling attacks can be crafted 
against the owners of these phone numbers by
exploiting cross-application features. 
Social and spear phishing attacks can be launched against 51,409 and 180,000 users respectively. Vishing attacks can be launched against 722,696 users. We also found 91,487 highly influential victims who can be attacked by crafting whaling attacks.
To evaluate the effectiveness of one of our attacks, phishing attacks, we conducted an online roleplay study with 314 participants on Amazon MTurk. 
Our results show that social phishing (69.2\%) is the most successful phishing attack on \ott, followed by spear (54.3\%), and non-targeted phishing (35.5\%). 
To mitigate the attacks demonstrated in this paper, we also suggest some recommendations for \ott and \caller which can prevent their users falling prey to targeted attacks.

\bibliographystyle{abbrv}
\balance
\bibliography{ref}
\section{Appendix} \label{sec:appendix}

\begin{figure*}
\subcapraggedrighttrue
\subfigure[Truecaller]
{ \label{fig:tc}
\includegraphics[width=3.21cm, trim=4.9cm 6.8cm 3.6cm 0cm, clip]{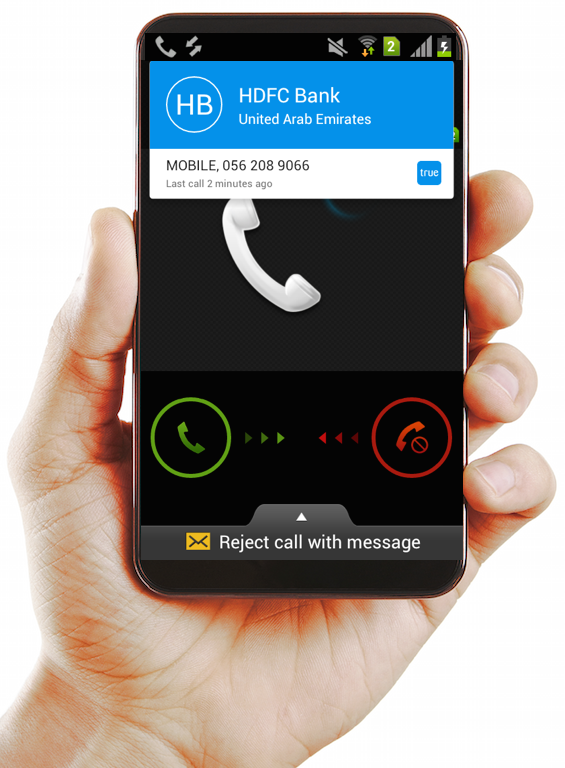}
}
\subfigure[Whitepages Caller ID \& Block]
{ \label{fig:whitepages}
\includegraphics[width=3.21cm, trim=4.9cm 6.8cm 3.6cm 0cm, clip]{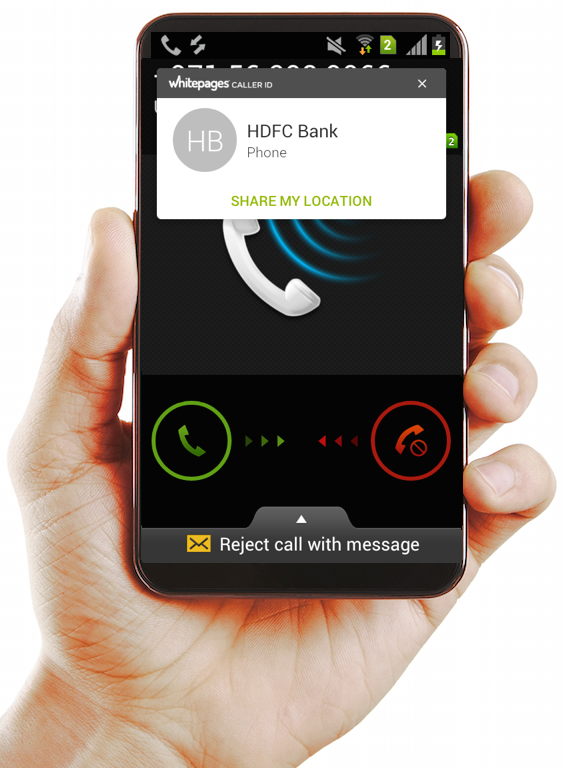}
}
\subfigure[Facebook's Hello]
{ \label{fig:hello}
\includegraphics[width=3.21cm, trim=4.9cm 6.8cm 3.6cm 0cm, clip]{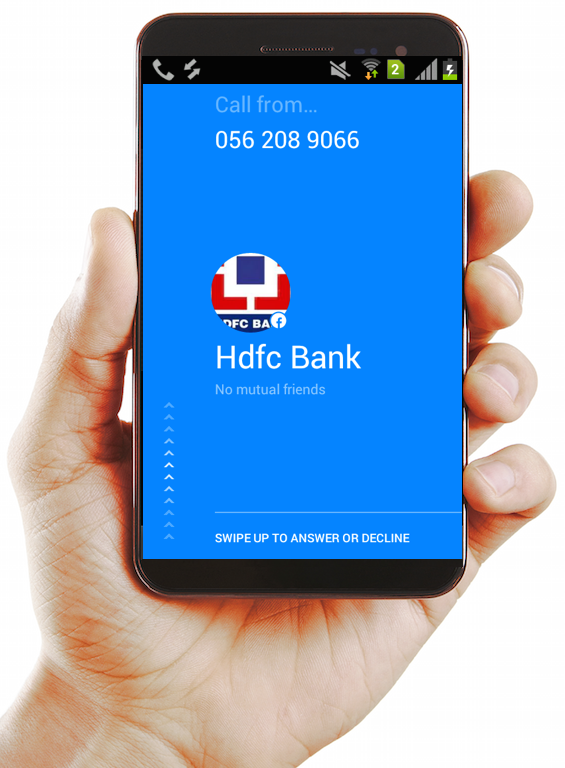}
}
\subfigure[Whoscall-Caller ID and Block]
{ \label{fig:whoscall}
\includegraphics[width=3.21cm, trim=4.9cm 6.8cm 3.6cm 0cm, clip]{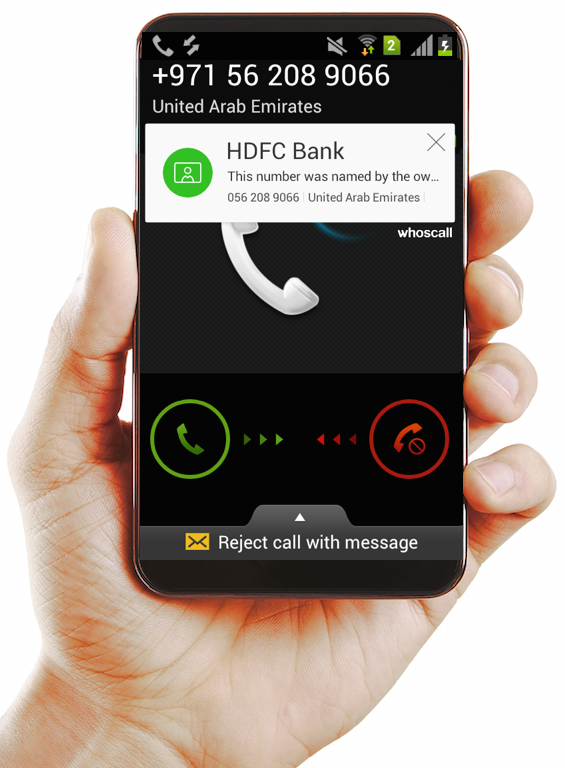}
}
\subfigure[Contactive]
{ \label{fig:contactive}
\includegraphics[width=3.21cm, trim=4.9cm 6.8cm 3.6cm 0cm, clip]{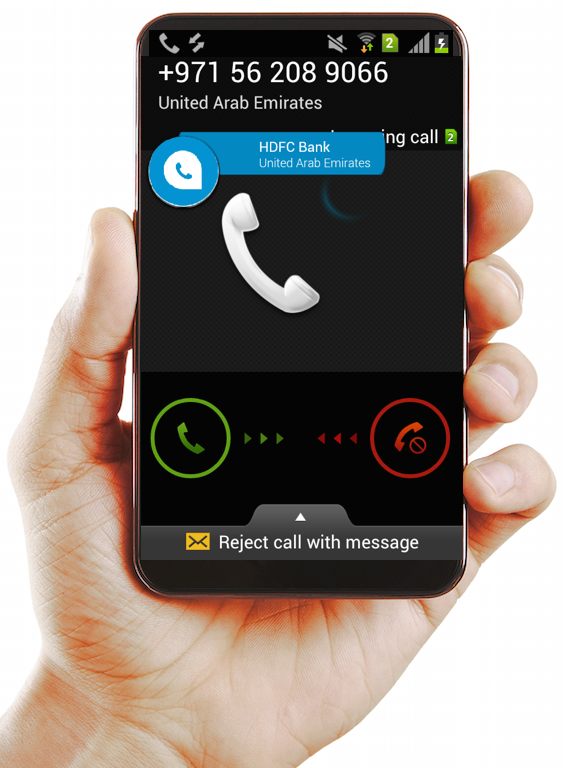}
}
\caption{Incoming call showing fake HDFC bank (example in our case) on various \caller.}
\label{fig:tc}
\end{figure*}

\end{document}